\documentclass[aps,prd,reprint,preprintnumbers,superscriptaddress,showpacs,twocolumn]{revtex4-1}
\usepackage{latexsym,graphicx,amssymb,amsmath,mathrsfs}
\usepackage{setspace,bm}
\usepackage[breaklinks, colorlinks=true, pdfstartview=FitV, linkcolor=red, citecolor=blue, urlcolor=blue]{hyperref}

\usepackage{amsmath}
\usepackage[usenames]{color}
\usepackage{latexsym}
\usepackage{epstopdf}

\newcommand\nc{N_\mathrm{c}}

\newcommand\mur{\mu_\mathrm{R}}
\newcommand\mui{\mu_\mathrm{I}}

\newcommand\qrw{\theta_\mathrm{RW}}
\usepackage[normalem]{ulem}  

\newcommand{\comment}[1]{}

\renewcommand\sout{\bgroup \color{red} \ULdepth=-.5ex \ULset}
\begin{document}
\preprint{YITP-17-60}
\preprint{RIKEN-QHP-313}

\title{Dirac-mode expansion of quark number density\\
 and its implications of the confinement-deconfinement transition}

\author{Takahiro M. Doi}
\email[]{takahiro.doi.gj@riken.jp}
\affiliation{Theoretical Research Division, Nishina Center, RIKEN, Wako 351-0198, Japan}

\author{Kouji Kashiwa}
\email[]{kouji.kashiwa@yukawa.kyoto-u.ac.jp}
\affiliation{Yukawa Institute for Theoretical Physics,
Kyoto University, Kyoto 606-8502, Japan}

\begin{abstract}
We investigate the quark number density at finite imaginary
 chemical potential by using the Dirac-mode expansion.
In the large quark mass region, it is found that the
 quark number density can be expressed by
 the Polyakov loop and its conjugate in all order of the large quark
 mass expansion.
Then, there are no specific Dirac-modes which dominantly contribute to the
 quark number density.
In comparison, the small quark mass region
 is explored by using the quenched lattice QCD simulation.
We found that the absolute value
 of the quark number density strongly depends on the low-lying
 Dirac-modes, but its sign does not.
 This means that the existence of the Roberge-Weiss transition which is
 characterized by the singular behavior of the quark number density is
 not sensitive to low-lying Dirac-modes.
This property enables us to discuss the confinement-deconfinement
 transition from the behavior of the quark number density via the quark
 number holonomy.
\end{abstract}

\maketitle

\def\slash#1{\not\!#1}
\def\slashb#1{\not\!\!#1}
\def\slashbb#1{\not\!\!\!#1}

\section{Introduction}

Understanding the phase structure of Quantum Chromodynamics (QCD) at
finite temperature ($T$) and real chemical potential $(\mur)$
is one of the interesting and important subjects in nuclear and
elementary particle physics.
The chiral and confinement-deconfinement transitions are key phenomena
for this purpose, but the confinement-deconfinement transition is not
yet fully understood comparing with the chiral transition.
Although the chiral transition can be described by the
spontaneous breaking of the chiral symmetry, but we can not
find any classical order-parameters of the confinement-deconfinement
transition in the system with dynamical quarks.

The Polyakov loop respecting the gauge-invariant holonomy becomes the
exact order-parameter of the spontaneous $\mathbb{Z}_{\nc}$ symmetry
breaking and then it has direct relation with the
confinement-deconfinement
transition in the infinite quark mass limit.
However, it is no longer the order parameter if the quark mass is
finite.
Other quantity such as the dual quark
condensate~\cite{Bilgici:2008qy,Fischer:2009wc,Kashiwa:2009ki,Benic:2013zaa,Xu:2011pz}
which has been proposed to characterize the confinement-deconfinement transition
still shares the same problem since it is based on the
spontaneous $\mathbb{Z}_{\nc}$ symmetry breaking.
Therefore, we need some extension of ordinary determinations to clearly
discuss and investigate the confinement-deconfinement transition in the
system with dynamical quarks.
One attempt is calculating the Polyakov-loop
fluctuation~\cite{Lo:2013hla,Doi:2015rsa} which is
originally introduce to attack the renormalization content of the
Polyakov-loop, but now it is considered as the good indicator of the
confinement-deconfinement transition comparing with the ordinary
analysis of the Polyakov loop.

Recently, the behavior of the quark number density at finite
imaginary chemical potential ($\mui$) has been used to determine the
confinement-deconfinement transition based on topological properties
of QCD.
In Refs.~\cite{Kashiwa:2015tna,Kashiwa:2016vrl,Kashiwa:2017yvy}, it has
been proposed that the topological change of QCD thermodynamics at
finite $\mui$ can be used to determine the confinement-deconfinement
transition.
This determination is based on the analogy of the topological order
discussed in the condensed matte
physics~\cite{Wen:1989iv} and QCD at $T=0$~\cite{Sato:2007xc}.
In discussions of the topological order at $T=0$, the
ground-state degeneracy plays a crucial role to determine
the topologically ordered and dis-ordered phases.
The nontrivial free-energy degeneracy in QCD induced by the
Roberge-Weiss (RW) transition at finite $\theta \equiv
\mui/T$~\cite{Roberge:1986mm}
can be considered as the analog of the ground-state
degeneracy at zero temperature~\cite{Kashiwa:2015tna}.

Based on the non-trivial free-energy degeneracy, the quark number
holonomy which is defined by the
contour integral of the quark-number susceptibility of
$\theta=0 \sim 2\pi$ has been proposed as the quantum order-parameter for the
confinement-deconfinement transition.
Since the quark number holonomy counts gapped points of the quark number
density along $\theta$ and thus it becomes non-zero (zero) in the
deconfined (confined) phase.
Particularly, the quark number density at $\theta=\pi/3$ is quite
important for the topologically determined confinement-deconfinement
transition because the RW transition is expected
to be happen here.
Below, we use the term {\it topological confinement-deconfinement
transition} when we determine the confinement-deconfinement transition by
using the quark number holonomy.
There are some papers which investigated the quark number density
at finite $\theta$~\cite{DElia:2002tig,DElia:2004ani,Takahashi:2014rta}, but
properties of the quark number density are not well understood yet.
Therefore, we investigate the quark number density and its implications
of the topological confinement-deconfinement transition in this article.

To investigate the quark number density and its implications of the
topological confinement-deconfinement transition, the Dirac-mode
expansion is a powerful and convenient tool.
The chiral condensate and also the Polyakov loop are already expressed
in terms of Dirac eigenvalues to investigate the relation
between the chiral symmetry breaking and
confinement~\cite{Gongyo:2012vx,Iritani:2013pga,Suganuma:2014wya}.
Then, we can see that dominant Dirac-modes in the chiral
condensate and Polyakov loop are quite different; the dominant modes are
low-lying Dirac-modes in the chiral condensate, but there are no specific modes
in the Polyakov loop in the case with light quark masses.
This fact may indicate that the behavior of dominant Dirac-modes can
used to clarify which quantities are sensitive to the
confinement-deconfinement transition.
Therefore, it is interesting to analyze which behaviors is realized in the
quark number density at finite $\theta$.

This paper is organized as follows.
In the next section, we show the the heavy quark mass
(hopping parameter) expansion of the quark number density.
The Dirac-mode expansion of the quark number
density are discussed in Sec.~\ref{Sec:DME}.
Then, we estimate the quark number density in the large and also small
quark mass regions by using the analytic calculation and the quenched
lattice QCD simulation, respectively.
Section \ref{Sec:summary} is devoted to summary and discussions.

\section{Heavy-mass expansion of quark number density}
\label{Sec:HQME}

In this study, we consider the ${\rm SU}(N_{\rm c})$ lattice QCD
on the standard square lattice.
We denote each sites as $x=(x_1,x_2,x_3,x_4) \ (x_\nu=1,2,\cdots,N_\nu)$
and link-variables as $U_\nu(x)$.
We impose the temporal periodic boundary condition for link-variables to
generate configurations in the quenched calculation to manifest the
imaginary-time formalism.

On the lattice, the quark number density is defined as
\begin{align}
n_q
 &=\frac{1}{V}\sum_x
   \Bigl\langle \bar{q}(x) \frac{\partial D}{\partial \mu}q(x) \Bigr\rangle
\nonumber\\
 &=\frac{1}{V}\left\langle {\rm Tr}_{\gamma, {\rm c}}
   \left[ \frac{\partial D}{\partial \mu} \frac{1}{D+m} \right] \right\rangle,
\label{DefQuarkNumber}
\end{align}
where ${\rm Tr}_{c,\gamma}\equiv \sum_x {\rm tr}_{\gamma, {\rm c}}$ denotes the functional trace
and ${\rm tr}_{\gamma, {\rm c}}$ is taken over spinor and color indices.
The operator $D$ is the Dirac operator and $m$ expresses the quark mass.
In this article, we use the Wilson-Dirac operator with the chemical
potential $\mu$ in the lattice unit as
\begin{align}
D
=&-\frac{1}{2}\sum_{k=1}^3
\left[P(+k)\hat{U}_k+P(-k)\hat{U}_{-k}\right] \nonumber \\
&-\frac{1}{2}\left[{\mathrm e}^\mu P(+4)\hat{U}_4 +
{\mathrm e}^{-\mu} P(-4)\hat{U}_{-4}\right] \nonumber \\
&+4\cdot \hat{1}
, \label{WilsonDiracOp}
\end{align}
where $\hat{1}$ is the identity matrix and
the link-variable operator $\hat{U}_{\pm\nu}$ is defined by the matrix element
\begin{align}
\langle
x | \hat{U}_{\pm\nu} |x' \rangle=U_{\pm\nu}(x)\delta_{x\pm\hat{\nu},x'},
 \label{LinkOp}
\end{align}
with $U_\nu\in {\rm SU}(N_{\rm c})$ and $P(\pm\nu)=1\mp\gamma_\nu$ with
$\nu=1,\cdots,4$. 
The derivative $\frac{\partial D}{\partial \mu}$ is explicitly written as
\begin{align}
\frac{\partial D}{\partial \mu}
=-\frac{1}{2}\left[{\mathrm e}^\mu P(+4)\hat{U}_4 -
{\mathrm e}^{-\mu} P(-4)\hat{U}_{-4}\right]. \label{DelMuWilsonDiracOp}
\end{align}
It should be noted that $\mu$ is the dimensionless chemical potential on
the lattice and it relates to $\theta$ as $\mathrm{Im}(\mu) = \theta/N_\tau$.

In the imaginary-time formalism,
the temporal anti-periodicity for $D$ should be imposed to manifest
the anti-periodic boundary condition of quarks.
To that end, we add a minus sign to the matrix element of
the temporal link-variable operator $\hat U_{\pm 4}$
at the temporal boundary of $x_4=N_4(=0)$:
\begin{eqnarray}
\langle {\bf x}, N_4|\hat U_4| {\bf x}, 1 \rangle
&=&-U_4({\bf x}, N_4),
\nonumber \\
\langle {\bf x}, 1|\hat U_{-4}| {\bf x}, N_4 \rangle
&=&-U_{-4}({\bf x}, 1)=-U_4^\dagger({\bf x}, N_t).
\label{eq:LVthermal}
\end{eqnarray}
In this notation, the Polyakov loop is expressed as
\begin{align}
L
&\equiv\frac{1}{N_c V}
 \sum_x {\rm tr}_c
 \Bigl\{\prod_{n=0}^{N_4-1} U_4(x+n\hat{4}) \Bigr\} \nonumber \\
&=-\frac{1}{N_{\rm c}V}{\rm Tr}_c \{\hat U_4^{N_4}\}.
\label{PolyakovOp}
\end{align}
The minus sign stems from the additional minus on $U_4({\bf s}, N_t)$
in Eq.(\ref{eq:LVthermal}).

\subsection{Leading-order contribution}
\begin{figure}[h]
\begin{center}
\includegraphics[scale=0.14]{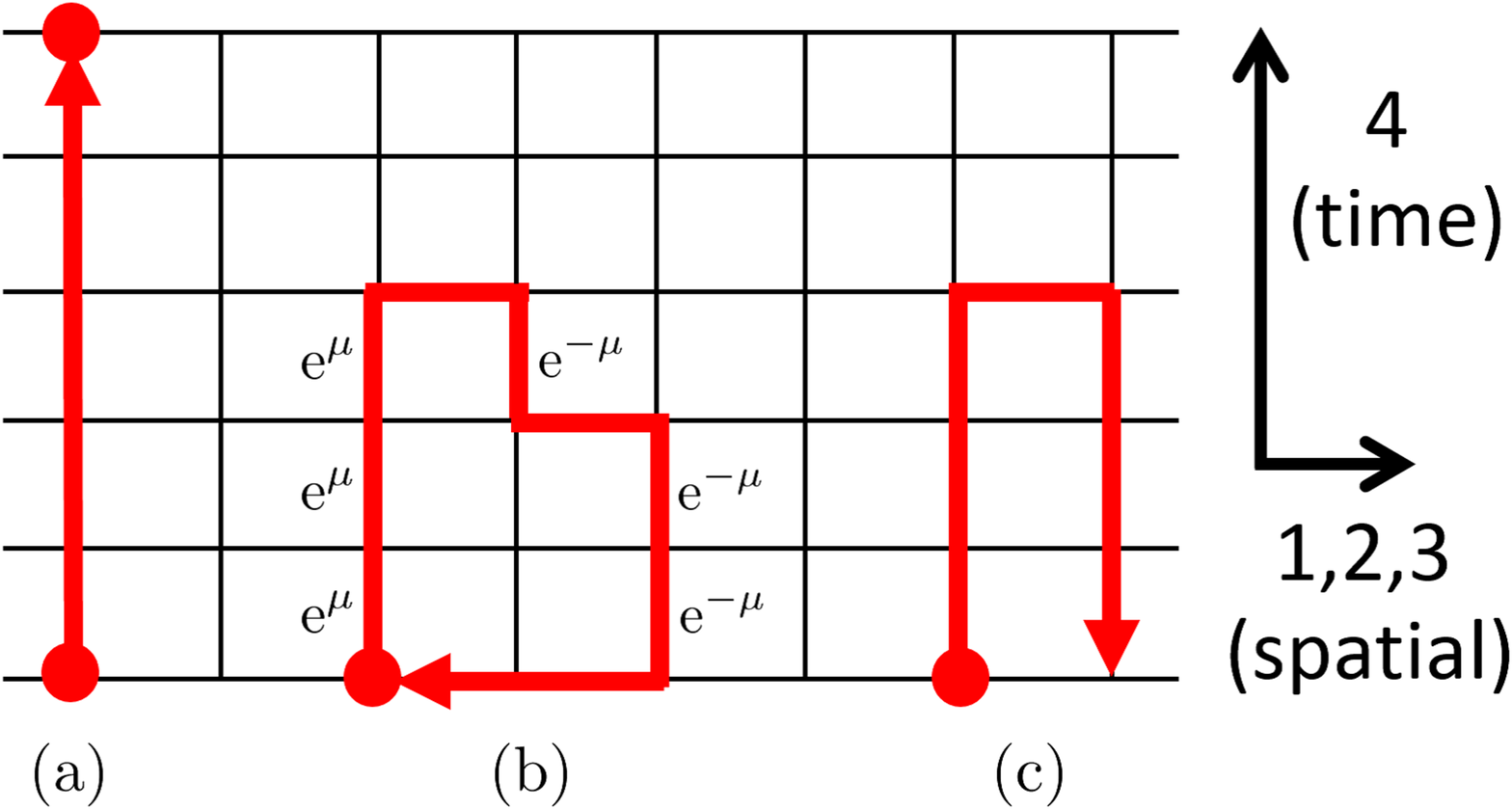}
\caption{
Several paths on the lattice in the case of $N_4=5$.
(a) A temporally closed loop. It is gauge-invariant and $\mu$-dependent.
(b) A spatially closed loop. It is gauge-invariant and $\mu$-independent.
(c) A non-closed loop. It is gauge-variant and thus it vanishes.
}
\label{loops}
\end{center}
\end{figure}

In the heavy quark mass region,
the quark number density (\ref{DefQuarkNumber}) can be expressed by
using the quark mass expansion as
\begin{align}
n_q
&=\frac{1}{MV}\left\langle {\rm Tr}_{\gamma, {\rm c}} \left[ \frac{\partial D}{\partial \mu}
\sum_{n=0}^\infty\left(-\frac{\hat{D}}{M}\right)^{n} \right] \right\rangle \\
&\equiv\frac{1}{MV}\sum_{n=0}^\infty \frac{c^{(n)}}{(-M)^n},
\label{QuarkNumberHeavyMass}
\end{align}
where
we define the effective mass $M\equiv m+4$ and the operator $\hat{D}\equiv D-4$. 
In Eq.~(\ref{QuarkNumberHeavyMass}), 
$c^{(n)}$ with smaller $n$ are relevant
because the effective mass $M$ is supposed to be large here.
Since the operator $\hat{D}$ and $\frac{\partial D}{\partial \mu}$ 
consists of linear terms in link-variables $U_\nu(x)$, 
the $n$-th order contribution $c^{(n)}$ can have many terms 
\begin{align}
c^{(n)}=c^{(n)}_1+c^{(n)}_2+c^{(n)}_3+\cdots, 
\label{n-thContribution}
\end{align}
where each $c^{(n)}_i$ is a product of $(n+1)$ link-variables; 
see Fig.~\ref{loops} as an example of paths on the lattice. 
It should be noted that many of them become exactly zero
because of Elitzur's theorem \cite{Elitzur:1975im};
only the gauge-invariant terms corresponding to closed loops are
nonzero.
Moreover, spatially closed loops which do not wind the temporal length 
are canceled out each other and have no contribution to the quark number density in total. 
Noting these important facts, it is confirmed that 
the $n$-th order contribution $c^{(n)}$ is constituted of 
the gauge-invariant loops with the length $(n+1)$ which winds the temporal direction. 
In particular, the nontrivial leading term in the expansion (\ref{QuarkNumberHeavyMass})
is the $(n=N_4-1)$-th order term $c^{(N_4-1)}$ 
which relates to the Polyakov loop ($L$) and its complex conjugate
(${\bar L}$).

Above result can be also understood from the $\mu$-dependence of the
quark number density.
Due to $n_q=0$ at $\mu=0$,
$\mu$-independent terms in the expansion (\ref{QuarkNumberHeavyMass})
must vanish.
As shown in Fig.~\ref{loops},
in the spatially closed loop,
the $\mu$-dependence are canceled out
between ${\mathrm e}^{\mu}U_4$ and ${\mathrm e}^{-\mu}U_4^\dagger$.
Thus, the nonzero terms must wind the temporal direction
using the temporal periodic boundary condition.
The other $\mu$-dependent terms up to $(n=N_4-1)$ order cannot form the
gauge-invariant closed-loop and
thus these should be vanished by Elitzur's theorem.
Therefore, the terms with $(n=N_4-1)$ which are proportional to $L$ and
${\bar L}$ are the dominant contributions for the quark number density
because $L$ is the shortest loop which leads to the $\mu$-dependence.

The actual contributions of the leading term, $c^{(N_4-1)}$, to the
quark number density at finite $\theta$ take the form;
\begin{align}
c^{(N_4-1)}\sim{\mathrm e}^{i\theta}L-{\mathrm e}^{-i\theta}L^*
&=2\sin(\theta+\phi)|L|.
\label{QuarkNumberLeading}
\end{align}
Since $\theta$, equivalently $N_\tau \hspace{1mm} \mathrm{Im}(\mu)$, can
be translated into the temporal
boundary condition of quarks,
$L$ and ${\bar L}$ can feel $\theta$-effects:
$L = |L| e^{i \phi}$ at $\theta = 2 \pi k$ exists in the trivial center
region ($\phi = 0$), but it can stay in the
non-trivial center region ($\phi \neq 0$)
at $\theta \neq 2 \pi k$ with $k \in \mathbb{Z}$.
For example, at sufficiently high $T$, we can obtain
$\phi=0$ in $0 < \theta < \pi/3$,
$\phi=4\pi/3$ in $\pi/3 < \theta < \pi$ and
$\phi=2\pi/3$ in $\pi < \theta < 5\pi/3$ because of the RW transition;
see Ref.~\cite{Roberge:1986mm} as an example.
The RW transition is also characterized by the gap of the
quark number density; the quark number density has opposite sign at
both side of the $\theta= \pi/3$ line.

In the heavy quark mass region,
the system is almost the quenched system and then we may
generate gauge configurations by only using the pure gauge action.
However, if we perfectly take the quenched limit,
the RW periodicity should vanish because
$\theta$-effects can not modify configurations.
Thus, we should take into account the discrete $\phi$-hopping,
$\phi=0 \rightarrow 4\pi/3\rightarrow 2\pi/3$ for
the variation of $\theta$ as a perturbation to reproduce the RW
periodicity and transition.
This treatment is implicitly used in the holographic QCD
calculations in the probe
limit~\cite{Aarts:2010ky,Rafferty:2011hd,Isono:2015uda}.
However, in the low $T$ region, the discrete $\phi$-hopping treatment
may provide wrong results since $L$ is smoothly rotating with varying of
$\theta$.
However, $|L|$ is exactly $0$ in the quenched calculation at low $T$
and thus there is no need to care the actual value of $\phi$.
In the following discussion, we consider the $\theta < \pi/3$ region and
thus the system is in the trivial center region.

\subsection{Higher-order contributions}
\begin{figure}[h]
\begin{center}
\includegraphics[scale=0.14]{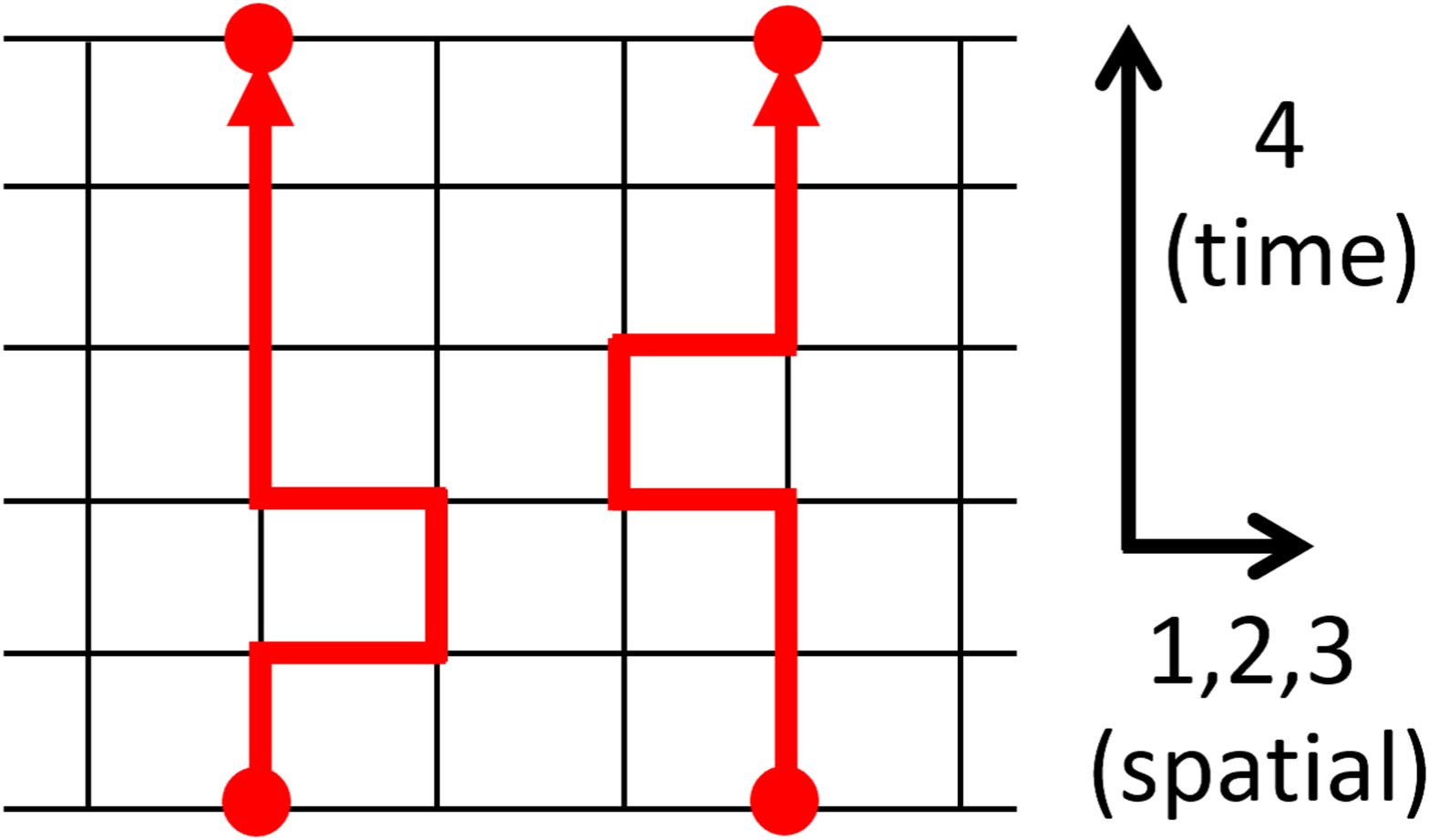}
\caption{
Examples of temporally closed loops corresponding to the sub-leading terms 
in the large-mass expansion (\ref{QuarkNumberHeavyMass}) 
in the case of $N_4=5$. 
(a) A temporally closed loop which makes a detour in the spatial direction. 
 It corresponds to Eq. (\ref{c1})
(b) A temporally closed loop which winds the temporal direction twice. 
 It corresponds to Eq. (\ref{c2})
}
\label{HigherOrder}
\end{center}
\end{figure}

In the heavy quark-mass expansion of the quark number
density~(\ref{QuarkNumberHeavyMass}),
there are higher order terms beyond the leading terms.
Noting again the important point in the previous subsection, 
each higher order term corresponds to 
a loop which winds the temporal direction and 
has the longer length $n>N_4$. 
Specifically, 
the sub-leading contributions are $(n=N_4+1)$ order's terms. 
For example, they include a term proportional to the quantity 
\begin{align}
c^{(N_4+1)}_1
\equiv
{\rm Tr}_c \{\hat U_4U_1U_4U_{-1}U_4^{N_4-2}\}.
\label{c1}
\end{align}
The sub-leading terms correspond to the closed paths 
which wind the temporal length and make a detour in the spatial direction. 
As another example, loops winding the temporal direction twice or more 
can be contribute to the expansion (\ref{QuarkNumberHeavyMass}). 
For example, a loop winding the temporal direction twice
\begin{align}
c^{(2N_4-1)}_1
\equiv
{\rm Tr}_c \{\hat U_4^{2N_4}\}
\label{c2}
\end{align}
is a possible contribution to Eq. (\ref{QuarkNumberHeavyMass}) 
as the ($n=2N_4-1$)-th order term. 
The examples of the higher order terms are shown in Fig.~\ref{HigherOrder}. 
In the next section, 
we show that 
all the terms in the heavy quark-mass expansion (\ref{QuarkNumberHeavyMass}) 
can be analytically represented in terms of the Dirac eigenmodes as well as the leading terms, 
namely the Polyakov loop and its conjugate. 

\section{Dirac-mode expansion of the quark number density}
\label{Sec:DME}
In this section,
we consider the quark number density in terms of the Dirac mode.
In large quark mass region,
we analytically show the statement
in all order of the large quark mass expansion (\ref{QuarkNumberHeavyMass}).
In small quark mass region,
we show that from numerical analysis by performing the quenched lattice
QCD simulation.

The Wilson-Dirac eigenvalues $\lambda_n$ are obtained from the
eigenvalue equation as
\begin{align}
D|n\rangle = \lambda_n|n\rangle, \label{DiracEigenEq}
\end{align}
where $|n\rangle$ is the Wilson-Dirac eigenstate.
For example, the chiral condensate $\langle \bar{q}q \rangle$ is defined as
\begin{align}
\langle \bar{q}q \rangle
=-\frac{1}{V}\left\langle {\rm Tr}_{\gamma, {\rm c}}
\left( \frac{1}{D+m} \right) \right\rangle,
\label{ChiralCondensate}
\end{align}
where $V$ is the four-dimensional volume.
Considering the Wilson-Dirac mode expansion of the chiral condensate, 
the low-lying eigenmodes of the operators $D$
have dominant contribution to the chiral condensate
known as Banks-Casher relation~\cite{Banks:1979yr, Giusti:2008vb}.

\subsection{large quark mass region}
We start the leading term to express it in terms of the Wilson-Dirac modes.
The leading contribution of the quark number density
in large quark mass region (\ref{QuarkNumberLeading}) is
expressed
by the Polyakov loop and its complex conjugate.
It is already known that the Polyakov loop can be expressed in terms of the
eigenmodes of the naive-Dirac operator which corresponds to the case of $r=0$
\cite{Suganuma:2014wya,Doi:2014zea}
and the Wilson-Dirac operator \cite{Suganuma:2016lnt,Suganuma:2016kva}.
In the following, we derive a different form of the Dirac spectral representation of the
the Polyakov loop using the operator $D$ on the square lattice
with the normal non-twisted periodic boundary condition for link-variables,
in both temporal and spatial directions.
We firstly define the following key quantity,
\begin{eqnarray}
I^{(N_4-1)}={\rm Tr}_{c,\gamma} (D\hat{U}_4^{N_4-1}).  \label{I0}
\end{eqnarray}
This quantity is defined by changing a temporal link-variable $\hat{U}_4$
of ${\rm Tr}_{c,\gamma}\hat{U}_4^{N_4}$
to the Wilson-Dirac operator $D$.
Substituting the definition (\ref{WilsonDiracOp}) of the Wilson-Dirac operator $D$,
the quantity $I^{(N_4-1)}$ can be calculated as
\begin{align}
I^{(N_4-1)}
&=-2\mathrm{e}^{-\mu}{\rm Tr}_{c,\gamma}(\hat{U}_4^{N_4})+({\rm other \ terms}) \nonumber \\
&=2\mathrm{e}^{-\mu}N_{\rm c}VL.  \label{I0_1}
\end{align}
Thus, the quantity $I^{(N_4-1)}$ is proportional to $L$.
Note that (other terms) vanish
because of the Elitzur's theorem and the trace over the Dirac indecies. 
On the other hand,
since $I^{(N_4-1)}$ in Eq. (\ref{I0}) is defined through the functional trace,
it can be expressed in the basis of Dirac eigenmodes as
\begin{align}
I^{(N_4-1)}
&=\sum_n\langle n|D\hat{U}_4^{N_4-1}|n\rangle + \mathcal{O}(a) \nonumber\\
&=\sum_n\Lambda_n \langle n|\hat{U}_4^{N_4-1}| n \rangle + \mathcal{O}(a).  \label{I0_2}
\end{align}
The $\mathcal{O}(a)$ term arises because 
Wilson-Dirac operator is not normal due to the $\mathcal{O}(a)$ Wilson term and 
the completeness of the Wilson-Dirac eigenstates has the $\mathcal{O}(a)$ error: 
\begin{align}
 \sum_n |n\rangle\langle n|=1+\mathcal{O}(a). 
 \label{Complete}
\end{align}
However, this error is controllable and can be ignored in close to the continuum limit. 

\begin{figure}[h]
\begin{center}
 \includegraphics[scale=0.15]{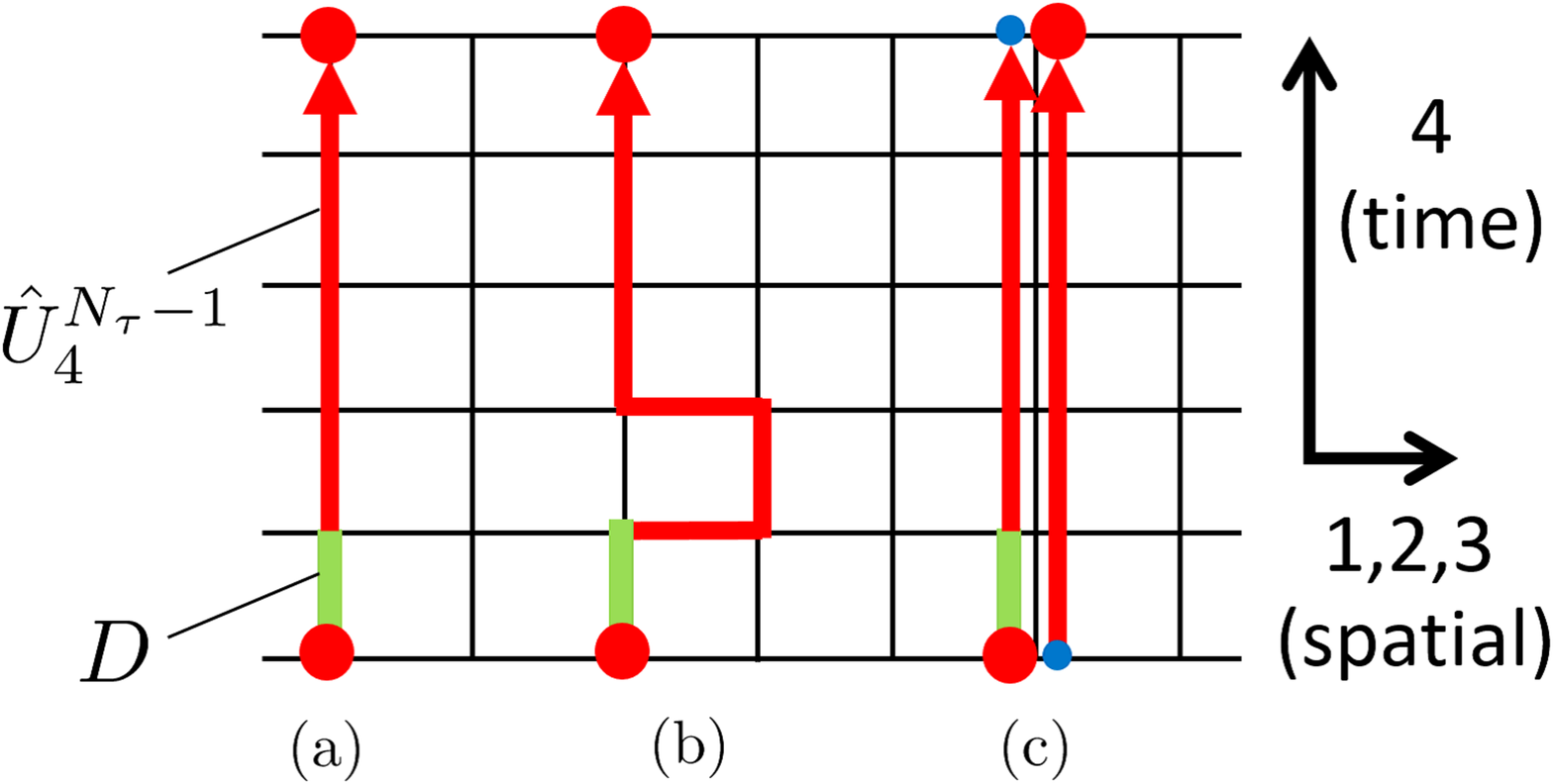}
\caption{
Schematic figures for the functional traces, Eqs. (\ref{I0}), (\ref{I1}) and (\ref{I2}). 
They are defined from the gauge-invariant quantities, 
 Eqs. (\ref{PolyakovOp}), (\ref{c1}) and (\ref{c2}), 
 by changing a temporal link-variable to the Wilson-Dirac operator. 
}
\label{Fig:ECQND}
\end{center}
\end{figure}

Combining Eqs. (\ref{I0_1}) and (\ref{I0_2}), 
one can derive the relation between $L$ and the Dirac modes as
\begin{eqnarray}
L\simeq\frac{\mathrm{e}^{-\mu}}{2N_{\rm c}V}
 \sum_n\Lambda_n\langle n|\hat{U}_4^{N_4-1}| n \rangle,
 \label{RelOrig}
\end{eqnarray}
with the $\mathcal{O}(a)$ error. 
This relation is a Dirac spectral representation of the Polyakov loop. 
From the formula (\ref{RelOrig}), 
it is analytically found that
the low-lying Dirac modes with $|\Lambda_n|\sim0$ have
negligible contribution to the Polyakov loop 
because the eigenvalue $\Lambda_n$ plays as the damping factor 
and the quantity $\langle n|\hat{U}_4^{N_4-1}| n \rangle$ 
has the finite value $|\langle n|\hat{U}_4^{N_4-1}| n \rangle|<1$.
It is also numerically shown that 
there is no dominant contribution in the Dirac modes 
to the Polyakov loop 
because the contributions of the Dirac modes whose eigenvalues are almost same 
are canceled due to the positive/negative symmetry 
of the Dirac-matrix element of the link-variable \cite{Doi:2014zea}. 
Thus, one can find the
Dirac spectrum representation of the quark
number density $n_q$ in the leading order 
and the low-lying Dirac modes 
have little contribution to the quark number density. 

The above discussion on the leading term can be applicable to 
the Dirac spectrum representation of the higher order terms. 
For example, a sub-leading term $c^{(N_4+1)_1}$ can be expressed as 
\begin{align}
c^{(N_4+1)}_1
\simeq
\frac{1}{2}
 \sum_n\Lambda_n\langle n|U_1U_4U_{-1}U_4^{N_4-2}| n \rangle,
 \label{c1_Dirac}
\end{align} 
by considering the functional trace 
\begin{eqnarray}
I^{(N_4+1)}\equiv{\rm Tr}_{c,\gamma} (DU_1U_4U_{-1}U_4^{N_4-2}),  \label{I1}
\end{eqnarray}
instead of Eq. (\ref{I0}). 
Moreover, the loop which winds the temporal direction twice can be expressed as 
\begin{align}
c^{(2N_4-1)}_1
\simeq
\frac{1}{2}
 \sum_n\Lambda_n\langle n|U_4^{2N_4-1}| n \rangle.
 \label{c2_Dirac}
\end{align} 
by considering the different functional trace 
\begin{eqnarray}
I^{(2N_4-1)}\equiv{\rm Tr}_{c,\gamma} (DU_4^{2N_4-1}).  \label{I2}
\end{eqnarray}
In the same way, all the terms in the expansion (\ref{QuarkNumberHeavyMass}) 
can be expressed in terms of the Wilson-Dirac modes. 
Thus, it is analytically found that 
the quark number density is insensitive to 
the density of the low-lying Wilson-Dirac modes in the all-order. 
However, this fact is only valid in the sufficiently large quark mass region
since other contributions which can not be expressed by $L$ and
${\bar L}$ can appear in the small $m$ region. 
Actually, every quark bilinears including the chiral condensate show the
same behavior of the quark number density in terms of Dirac modes 
in the large quark mass region.
However, we already know that the dominant contributions of the chiral
condensate are low-lying Dirac modes in the small quark mass region.

\subsection{Small quark mass region}
In the small quark mass region, the quark number density
requires contributions which can not be expressed by $L$ and ${\bar L}$
and thus we perform the lattice QCD simulation to investigate the quark
number density.
In this study, we perform the quenched calculation 
with the ordinary plaquette action 
and then fermionic
observables are evaluated by using the Wilson-Dirac operator (\ref{WilsonDiracOp}) 
with the imaginary chemical potential $\mu$. 
Our calculation is performed in both the confinement phase and 
the deconfinement phase. 
In the confinement phase, 
we consider $6^4$ lattice with $\beta\equiv 6/g^2=5.6$ and $\mu=(0,1745)$ 
which corresponds to $a\simeq0.25$ fm and $T\simeq133$ MeV.
In the deconfinement phase, 
we consider $6^3\times5$ lattice with $\beta=6.0$ and $\mu=(0,2094)$ 
which corresponds to $a\simeq0.10$ fm and $T\simeq400$ MeV.
Both values of $\mu$ correspond to $\theta\simeq\pi/3$. 
In both cases, we set the quark mass as $m=-0.7$ in the lattice unit, 
which is equivalent to the hopping parameter $\kappa\equiv1/(2m+8)\simeq0.151515$, 
for the calculation of the eigenmodes of the Wilson-Dirac operator 
in the small quark mass region \cite{Aoki:1999yr}.

In the lattice QCD simulation, there is the smearing of the phase
transition because of the finite size effect.
Of course, the RW transition which plays a crucial role
in the behavior of the quark number holonomy should be smeared in the
system with dynamical quarks.
However, in the quenched lattice QCD simulation, the RW transition
can reproduced by considering the discrete $\phi$-hopping and then the
smearing of the phase transition does not matter.
In addition, we can clarify the critical temperature of the
confinement-deconfinement transition from
previous quenched lattice QCD simulations because the RW endpoint and
the ordinary critical temperature determined by the Polyakov loop are
consistent
with each other in the quenched lattice QCD simulation.
This knowledge enable us to
safely clarify which temperature is above the critical temperature
when we investigate the quark number density.

The eigenstates of the Wilson-Dirac operator do not form the complete system,
and then we calculate the quark number density as
\begin{align}
\langle n_q \rangle
&=\frac{1}{2V}\left\langle {\rm Tr}_{\gamma, {\rm c}}
\left[
\frac{\partial D}{\partial \mu} \frac{1}{D+m}
-\left(\frac{\partial D}{\partial \mu} \frac{\gamma_4}{D+m} \right)^\dagger
\right] \right\rangle \nonumber \\
&\simeq\frac{i}{V}{\rm Im}\left\langle \sum_n
\Bigl\langle n \Bigl| \frac{\partial D}{\partial \mu} \Bigr|n \Bigr\rangle \frac{1}{\Lambda_n+m} \right\rangle.
\label{QuarkNumber_Dirac}
\end{align}
This form trivially takes pure imaginary value up to the
$\mathcal{O}(a)$ error.
Each contribution, $n_q^n$, to the quark number density 
of the Dirac mode with $\Lambda_n$ 
can be defined as
\begin{align}
n_q^n=
\frac{i}{V}{\rm Im} \Bigl \langle n \Bigl| \frac{\partial D}{\partial \mu} \Bigr|n \Bigr\rangle \frac{1}{\Lambda_n+m},
\end{align}
and then the quark number density becomes
\begin{align}
 n_q=\sum_n n_q^n.
\end{align}

Top (bottom) panel of Fig.~\ref{Fig:ECQND} shows each Dirac-mode contribution
to the quark number density, ${n}_\mathrm{q}^n (\Lambda)$, with
$\mu = (0,0.1745)$ and $\mu =(0,0.2094)$.
We here only show results with one particular configuration.
\begin{figure}[h]
\begin{center}
 \includegraphics[scale=0.65]{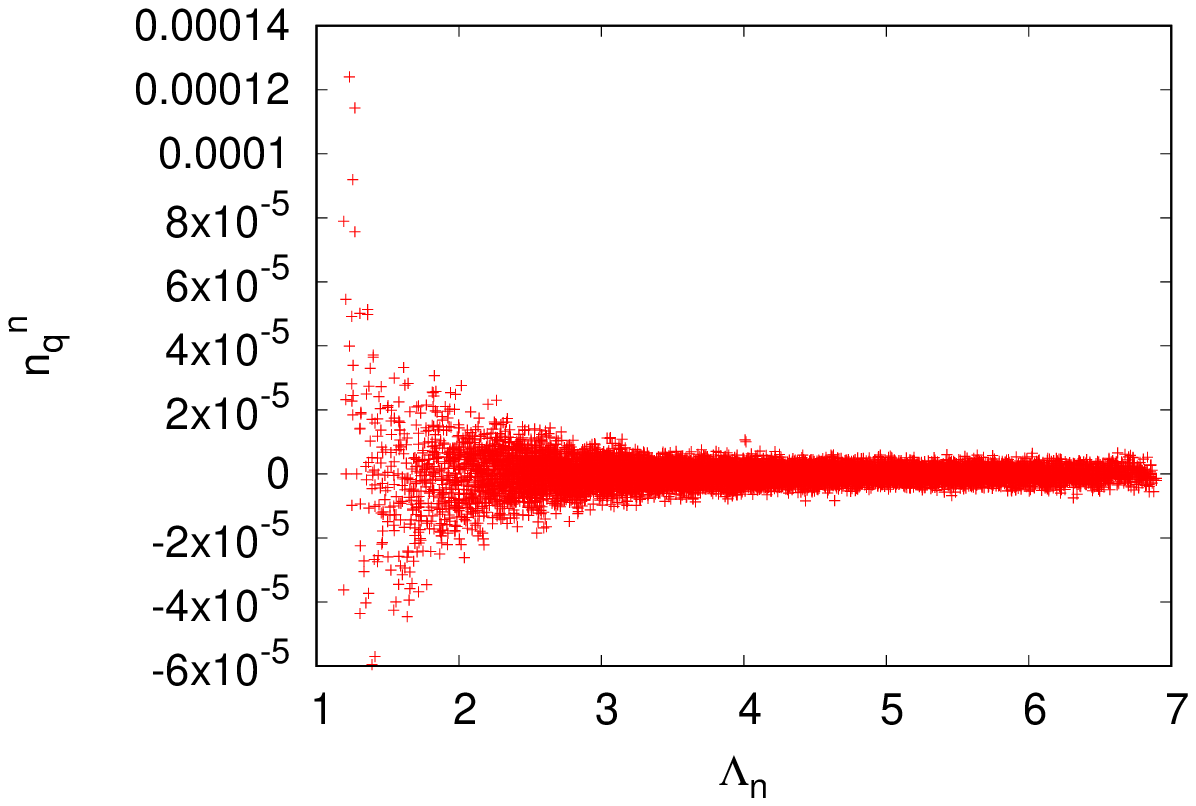}
 \includegraphics[scale=0.65]{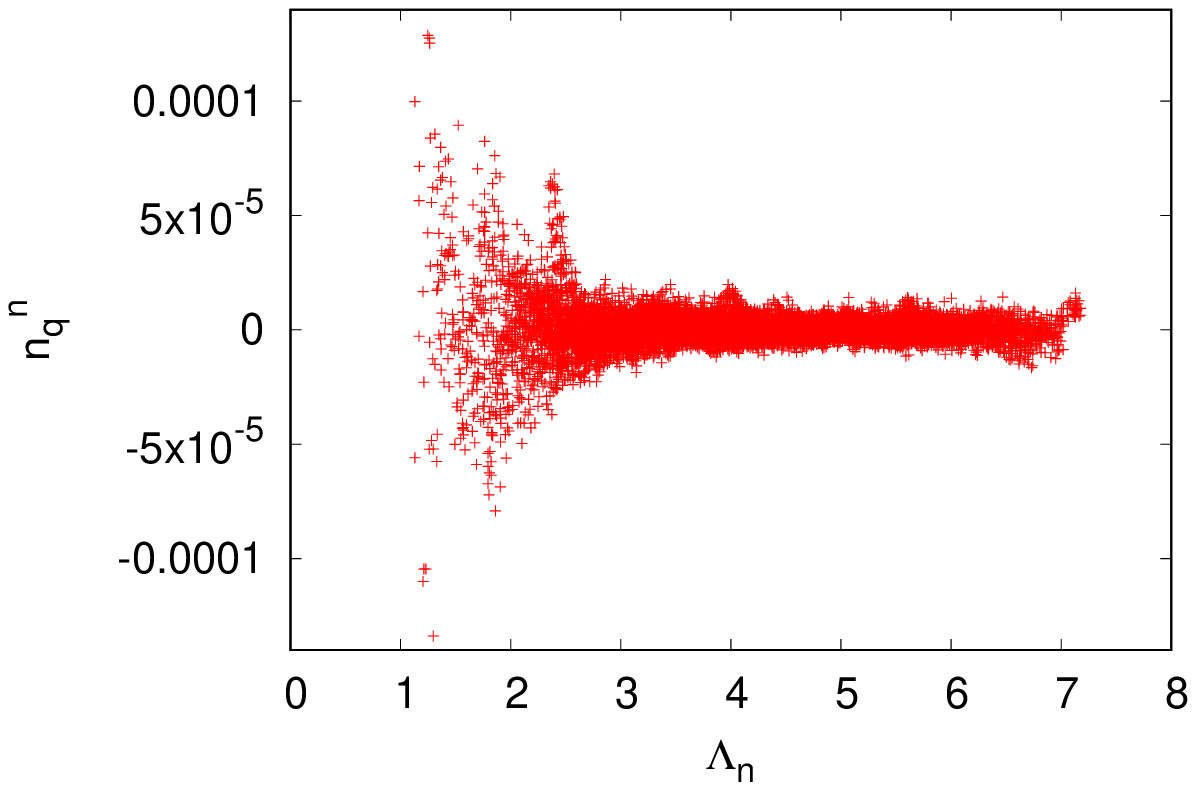}
\caption{
 Each Dirac-mode contribution to the quark number density as a function
 plotted against the Wilson-Dirac eigenvalue $\Lambda_n$.
 Top and bottom panels show
 the result with $\mu =(0,0.1745)$ and $\mu =(0,0.2094)$, respectively.
 }
\label{Fig:ECQND}
\end{center}
\end{figure}
By comparing with both figures,
in the distribution of $n_\mathrm{q}$ at $\mu =(0,0.1745)$,
the positive-negative symmetry~\cite{Doi:2014zea} is almost realized,
which was originally found in the confined phase at $\mu=0$.
On the other hand, in the case of $\mu =(0,0.2094)$,
the symmetry is broken.
This means that the violation of the positive-negative symmetry leads to
$n_\mathrm{q} \neq 0$ at finite $\theta$.

\begin{figure}[t]
\begin{center}
 \includegraphics[scale=0.65]{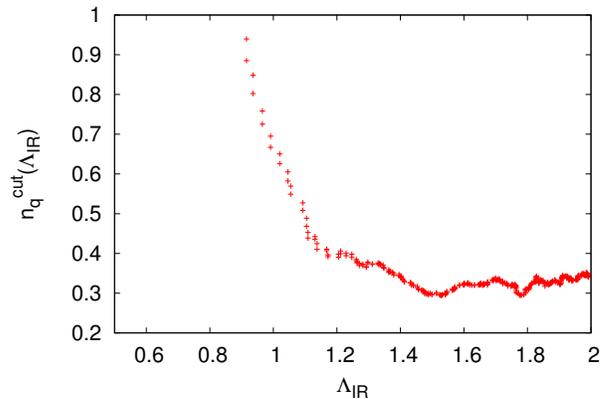}
\caption{
 The $\Lambda_\mathrm{cut}$-dependence of $n_\mathrm{q}^\mathrm{cut}$
 at $\mu =(0,0.2094)$.
}
\label{Fig:CQND}
\end{center}
\end{figure}

To investigate the quark number density in terms of Dirac modes,
we show the infra-red (IR) cutted quark number density with the cutoff
$\Lambda_\mathrm{cut}$ defined by
\begin{align}
 n^\mathrm{cut}_\mathrm{q} (\Lambda_\mathrm{IR})
 = \frac{1}{n_\mathrm{q}} \sum_{|\Lambda_n|>\Lambda_{\rm IR}}
 {n}_\mathrm{q}^n.
 \label{Eq:IR}
\end{align}
at $\mu = (0,0.2094)$ in Fig.~\ref{Fig:CQND}.
In the evaluation of Eq.~(\ref{Eq:IR}), the configuration averaging is
basically possible but it misses a physical meaning because the averaging
well works after summing over all Dirac-modes.
Thus, we here show
$n^\mathrm{cut}_\mathrm{q}$ in one particular configuration.
We can see that the absolute value of the quark number density seems to
depend on the low-lying Dirac modes, but its sign does not.
This tendency can be found in almost all our configurations.
It means that the absolute value of the quark number density
shares a same property in terms of
Dirac modes with the chiral condensate, while its sign shares the
property with the Polyakov loop.

Finally, we discuss the topological confinement-deconfinement
transition from Dirac-mode analysis.
The order-parameter of the topological confinement-deconfinement
transition can be expressed~\cite{Kashiwa:2016vrl} as
\begin{align}
\Psi &=
   \oint_{0}^{2\pi}
   \Bigl\{\mathrm{Im} \Bigl(
          \frac{d {\tilde n}_q}{d \theta} \Bigl|_T \Bigr) \Bigr\}
         ~d\theta,
 \label{Eq:psi}
\end{align}
where ${\tilde n}$ is the dimensionless quark number density such as
${\tilde n}_\mathrm{q} = n_\mathrm{q}/T^3$.
It counts gapped points of the quark number density along
$\theta$ direction and thus it becomes zero (non-zero) in the confined
(deconfined) phase.
Equation (\ref{Eq:psi}) can be expressed as
\begin{align}
\Psi &= \pm
   2 \nc \lim_{\epsilon \to 0}
   \Bigl[ \mathrm{Im}~
 {\tilde n}_q ( \theta=\qrw \mp \epsilon )\Bigr],
\label{psi1}
\end{align}
when the RW endpoint which is the endpoint of the RW transition line
becomes the second-order point at
$\theta_\mathrm{RW} =\pi / 3$.
In Eq.~(\ref{psi1}), $\lim_{\epsilon \to 0} n_\mathrm{q} (\qrw \mp
\epsilon)$ characterizes
$\Psi$ and thus $\Psi$ shares the same property about the Dirac modes
with $n_\mathrm{q}$.
The important point is that the absolute value of $n_q$ does not have
so much meaning even if it is non-zero, but its sign is important since
the sign flipping at $\theta = (2k-1) \pi/3$
characterizes the gapped points along $\theta$-direction.
From our quenched lattice QCD data, the sign of the quark number density
are insensitive to the low-lying Dirac-modes and this behavior is similar
to the Polyakov loop. 
Equation (\ref{RelOrig}) holds 
also in the small quark mass regime 
because its derivation does not depend on the quark mass.
It should be noted that recent lattice QCD
data~\cite{D'Elia:2009qz,Bonati:2010gi} predict that the RW
endpoint seems to be the
triple-point where three first-order transition lines meet.
We should modify the expression (\ref{psi1}) in this case, but
we can expect that the dependence of
the Dirac modes is same with that of the second-order RW endpoint
scenario; $\Psi$ can be expressed by using
$n_\mathrm{q}(\theta_\mathrm{Beard})$ where $\theta_\mathrm{Beard}$ is
the endpoint value of $\theta$ of the
triple line at $\theta \neq \qrw$.

The $2+1$ flavor lattice QCD simulation
predicts that the RW endpoint is $208(5)$ MeV~\cite{Bonati:2016pwz} and
thus it becomes the critical temperature of the topological
confinement-deconfinement transition when the RW endpoint is second order.
On the other hand, the chiral pseudo critical temperature is about $155$
MeV~\cite{Aoki:2006br,Aoki:2009sc,Borsanyi:2010bp,Bazavov:2011nk}.
This result strongly supports that the topologically determined
confinement-deconfinement transition does not have the exact one-to-one
correspondence with the chiral phase transition.

\section{Summary and discussion}
\label{Sec:summary}

In this paper, we have discussed properties of the quark number
density at finite temperature ($T$) and imaginary chemical potential
($\mu_\mathrm{I}$) in terms of Dirac-modes.
In the heavy quark mass region, we use the heavy quark mass expansion
and then the analytic form is discussed.
On the other hand, we employ the quenched lattice QCD simulation in the
small quark mass region.

From the heavy quark mass expansion with the Dirac mode expansion,
we found that low-lying Dirac modes do not dominantly contribute to the
quark number density in all order of the heavy quark mass expansion.
Some other quark bilinears should also be insensitive to low-lying Dirac
modes. This result is valid if the quark number density can be well
expressed by the Polyakov loop ($L$) and its conjugate $({\bar L})$.
If some other contributions which can not be expressed by
$L$ and ${\bar L}$ appear, low-lying Dirac modes can become dominant modes.

The small quark mass region has been explored by
using the quenched lattice QCD simulation.
We found that the absolute value of the quark number density strongly
depends on low-lying Dirac modes, but its sign does not.
Therefore, the sign of the quark number density shares similar
properties with the Polyakov loop in terms of Dirac eigenvalues.
This result means that the RW transition at $\theta=\pi/3$ is
insensitive to low-lying Dirac modes where $\theta=\mu_\mathrm{I}/T$.

The RW transition plays a crucial role in the determination of the topological
confinement-deconfinement transition and thus we can discuss the
transition from the viewpoint of Dirac eigenvalues.
The order-parameter of the topological confinement-deconfinement
transition is the quark number holonomy ($\Psi$) which is defined by the
contour integral of the quark number susceptibility along $\theta=0 \sim
2\pi$.
This quantity becomes non-zero if the quark number density has the gap
along $\theta$ with fixed $T$ and then the deconfined phase is realized.
From our quenched lattice QCD data, it is found that
the quark number holonomy is sensitive to the confinement properties of
QCD because the sign of the quark number density shares similar
properties with the Polyakov loop.
Therefore, our results support that the quark number holonomy is
the good quantum order parameter for the confinement-deconfinement
transition.

It is interesting to compare the quark number holonomy with other
quantities which relates with the confinement-deconfinement transition.
The dual quark condensate defined with the twisted boundary condition
is one of sensitive probes for the quark-deconfinement
\cite{Bilgici:2008qy,Fischer:2009wc,Kashiwa:2009ki,Benic:2013zaa,Xu:2011pz}.
In the calculation of the dual quark condensate, we
must break the RW periodicity because it should be zero if the RW
periodicity exists.
It is usually done by imposing the twisted boundary condition on the
Dirac operator, while configurations are created under the anti-periodic
boundary condition in the system with dynamical quarks.
However, this procedure is not unique.
Therefore, there is the uncertainty in the determination of the dual
quark condensate.
Also, it is well known that this quantity is strongly affected by other
phase transitions~\cite{Benic:2013zaa,Marquez:2015bca,Zhang:2015baa}.
Another famous one is the QCD monopole.
The QCD monopole which is also the order-parameter of the chiral
symmetry appears by fixing the maximally Abelian gauge
and plays important role
in the dual-superconductor picture for the mechanism of confinement
\cite{Suganuma:1993ps}.
In fact, after removal of the QCD monopole
from the QCD vacuum generated in the lattice QCD,
both chiral restoration and 
quark deconfinement occur \cite{Miyamura:1995xn,Woloshyn:1994rv}.
The quark number holonomy has advantages over
the QCD monopole and the dual quark condensate
in the view point of the gauge invariance and the temporal boundary condition.
While the QCD monopole is gauge-variant,
the quark number holonomy is gauge-invariantly defined.
Also, on the one hand,
the dual quark condensate can be defined with only the twisted boundary condition.
On the other hand,
the quark number holonomy can be defined with arbitrary boundary condition,
including the periodic boundary condition, which is needed for finite-temperature formalism.
Also, the dual quark condensate has uncertainty in the actual
calculation process,
but the quark number holonomy does not. 
Thus, the quark number holonomy
is superior order parameter of the confinement-deconfinement transition.
Moreover, the quark number holonomy is better than the Polyakov loop
because the Polyakov loop works as the order parameter only at the large
quark-mass regime.

\vspace{2mm}
{\it Acknowledgments:}
T.M.D. is supported by 
the Grantin-Aid for JSPS fellows (No.15J02108) 
and the RIKEN Special Postdoctoral Researchers Program.

\bibliography{ref.bib}

\end{document}